\documentclass[aip,rsi,reprint,graphicx]{revtex4-1}
\usepackage{graphicx}
\usepackage{color}

\usepackage[section]{placeins}
\usepackage{placeins}
\usepackage[nice,loose]{units}

\begin{document}

\title{Confocal shift interferometry of coherent emission from trapped dipolar excitons}

\author{J. Repp}
\affiliation{Walter Schottky Institut and Physik-Department, Am Coulombwall 4a, Technische Universit\"at M\"unchen, D-85748 Garching, Germany}
\affiliation{Nanosystems Initiative Munich (NIM), Schellingstr. 4, 80799 M\"unchen, Germany.}
\affiliation{Center for NanoScience and Fakult\"at f\"ur Physik,
Ludwig-Maximilians-Universit\"at,
Geschwister-Scholl-Platz 1, 80539 M\"unchen, Germany}

\author{G. J. Schinner}
\affiliation{Nanosystems Initiative Munich (NIM), Schellingstr. 4, 80799 M\"unchen, Germany.}
\affiliation{Center for NanoScience and Fakult\"at f\"ur Physik,
Ludwig-Maximilians-Universit\"at,
Geschwister-Scholl-Platz 1, 80539 M\"unchen, Germany}

\author{E. Schubert}
\affiliation{Nanosystems Initiative Munich (NIM), Schellingstr. 4, 80799 M\"unchen, Germany.}
\affiliation{Center for NanoScience and Fakult\"at f\"ur Physik,
Ludwig-Maximilians-Universit\"at,
Geschwister-Scholl-Platz 1, 80539 M\"unchen, Germany}

\author{A. K. Rai}
\affiliation{Angewandte Festk\"orperphysik, Ruhr-Universit\"at Bochum, Universit\"atsstra{\ss}e 150, 44780 Bochum, Germany}

\author{D. Reuter}
\affiliation{Angewandte Festk\"orperphysik, Ruhr-Universit\"at Bochum, Universit\"atsstra{\ss}e 150, 44780 Bochum, Germany}
\affiliation{Universit\"at Paderborn, Department Physik, 33098 Paderborn, Germany}

\author{A. D. Wieck}
\affiliation{Angewandte Festk\"orperphysik, Ruhr-Universit\"at Bochum, Universit\"atsstra{\ss}e 150, 44780 Bochum, Germany}

\author{U. Wurstbauer}
\affiliation{Walter Schottky Institut and Physik-Department, Am Coulombwall 4a, Technische Universit\"at M\"unchen, D-85748 Garching, Germany}
\affiliation{Nanosystems Initiative Munich (NIM), Schellingstr. 4, 80799 M\"unchen, Germany.}

\author{J. P. Kotthaus}
\affiliation{Center for NanoScience and Fakult\"at f\"ur Physik,
Ludwig-Maximilians-Universit\"at,
Geschwister-Scholl-Platz 1, 80539 M\"unchen, Germany}

\author{ A. W. Holleitner}
\affiliation{Walter Schottky Institut and Physik-Department, Am Coulombwall 4a, Technische Universit\"at M\"unchen, D-85748 Garching, Germany}
\affiliation{Nanosystems Initiative Munich (NIM), Schellingstr. 4, 80799 M\"unchen, Germany.}
\date{\today}

\begin{abstract}
We introduce a confocal shift-interferometer based on optical fibers. The presented spectroscopy allows measuring coherence maps of luminescent samples with a high spatial resolution even at cryogenic temperatures. We apply the spectroscopy onto electrostatically trapped, dipolar excitons in a semiconductor double quantum well. We find that the measured spatial coherence length of the excitonic emission coincides with the point spread function of the confocal setup. The results are consistent with a temporal coherence of the excitonic emission down to temperatures of 250 mK.
\end{abstract}

%
\maketitle
In 1962 a Bose-Einstein-condensation of excitons was predicted by Blatt \cite{Blatt1962} and Moskalenko \cite{Moskalenko1962}. Dipolar excitons, where the composing electrons and holes are located in adjacent quantum wells of a semiconductor hetero-structure \cite{Gaertner2006,Andreakou2014}, are particularly suited to probe such a condensate. Due to the reduced overlap of the electron and hole wave functions, the lifetime of these quasi-particles is strongly increased. This enables them to cool down to the lattice temperature and to relax to their quantum mechanical ground state. To reach a Bose-Einstein condensate, a high particle density and low temperatures are a prerequisite, as demonstrated for related quasi-particles, namely polaritons \cite{Kasprzak2006}. For dipolar excitons, a condensation has been recently reported \cite{High2012_2,Shilo2013,Alloing2014}. In particular, shift-interferometry was used to study the spatial coherence of the excitonic emission which is considered as an indication of a condensed excitonic phase \cite{High2012_2,Alloing2014,Rossier1998,Laikhtman1998}. Yet, there has been a controversial discussion about the impact of a limited spatial resolution on the ability to resolve the spatial coherence of the excitonic emission \cite{Semkat2012,Snoke2011}. \newline
Here, we introduce a fiber-based confocal interferometer which operates down to cryogenic temperatures. We demonstrate that the approach allows measuring the coherence variation of the photoluminescence (PL) of trapped, dipolar excitons with a high spatial resolution. We extract a coherence length on the micrometer scale which is consistent with earlier reports\cite{Alloing2014}. However, we also observe that this length does not exceed the point spread function of our confocal set-up with two objectives even at the lowest temperature of 250 mK. In turn, we interpret the data such that a temporal coherence dominates the excitonic emission in our samples. \newline
\begin{figure}
\begin{center}
\includegraphics[width=8.5cm]{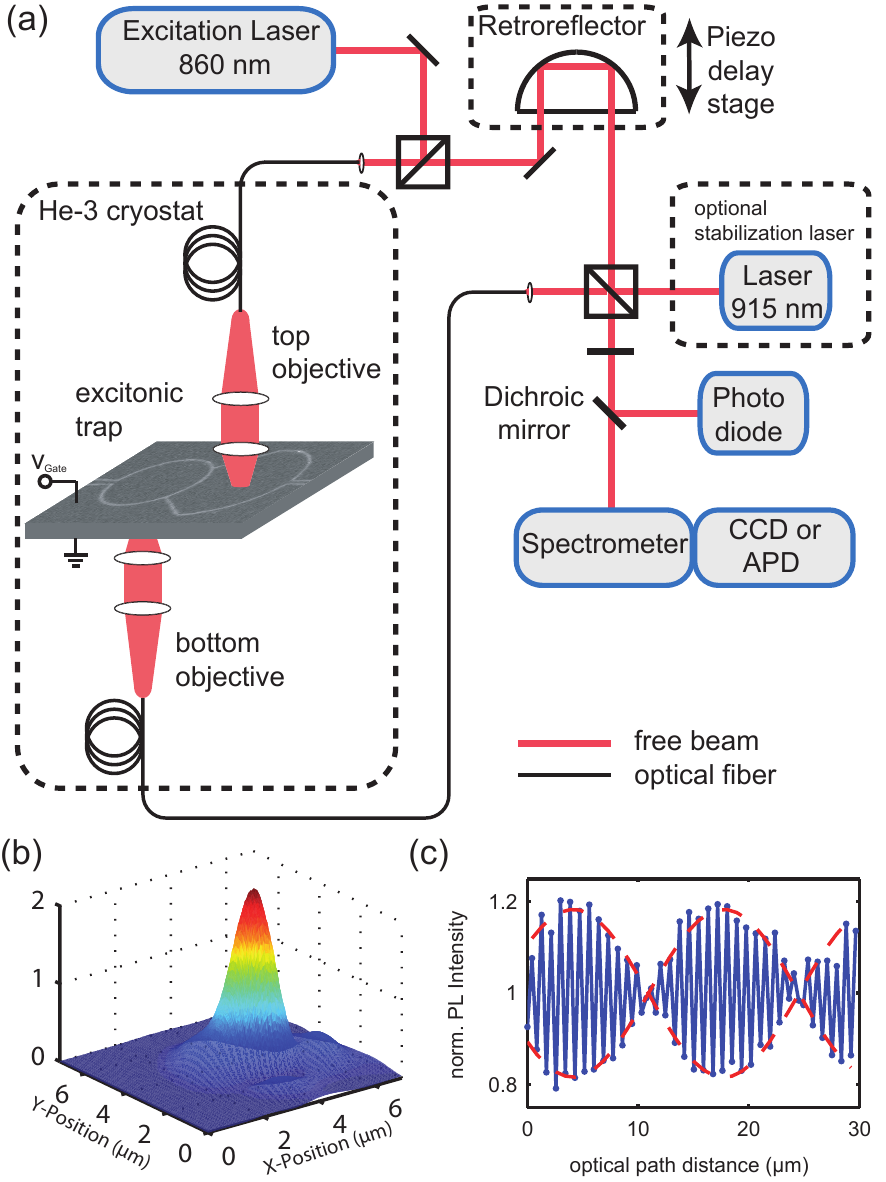}
\end{center}
\caption{\label{fig1} Color online (a) Sketch of the fiber-based interferometer with two independent confocal objectives operated in a He-3 cryostat. (b) Spatial intensity map of the light transmitted from the top to the bottom objective at a laser wavelength of $\lambda_{\text{laser}} = 860$ nm. (c) Interferogram recorded at the position, where both foci are at the same position. Dashed curve is a fit according to eq.(1).
}
\end{figure}
We begin with the interferometer. It comprises two confocal objectives which can be positioned individually by piezo-positioners [Fig. 1(a)]\cite{Schinner2013_2}. The top objective is used to optically excite the sample with a pulsed laser.  
The sample\textquoteright s PL is collected with both objectives and accordingly coupled into the two arms of the fiber-based interferometer. The interference signal is detected by a charge coupled device and/or an avalanche photodiode for time-resolved measurements. To measure the point spread function (PSF) of the optical system, the fraction of laser light is detected which is transmitted from the top to the bottom objective while the top objective is kept fixed and the bottom objective is scanned across the sample [Fig. 1(b)]. The full width at half maximum (FWHM) of the transmitted signal is $1.4 \pm 0.1$ $\mu$m at the used laser wavelength of $\lambda_{\text{laser}} = 860 $ nm. This definition of the PSF of the overall setup corresponds to the convolution of the PSFs of each individual objective; giving a factor of $\sqrt{2} $ \cite{Semkat2012}.We independently determine the spatial resolution of the setup to be $d_{\text{resolution}} = 935 \pm 65 $ nm, which compares well with the diffraction-limited value $\lambda_{\text{laser}}/(2\cdot\text{NA})$ with the given numerical aperture NA$ = 0.68$ of the objectives. Convoluting the resolution of both objectives gives a Gaussian with a FWHM of $1.3 \pm 0.1$ $\mu$m. This value agrees with the PSF as defined above. \newline
\begin{figure}
\centering
\includegraphics[width=8.5cm]{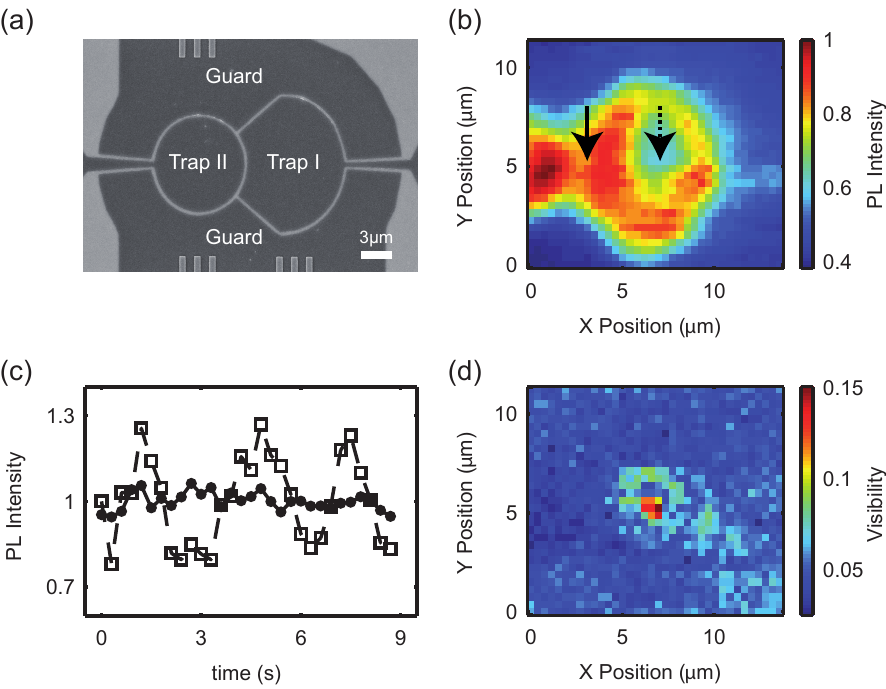}
\caption{\label{Fig2} Color online (a) Scanning electron microscope image of the top gates (dark gray) of the sample; i.e. Trap I, Trap II, and Guard. (b) Sum of the PL intensities of both objectives. (c) Interferogram of the PL of dipolar excitons. Top objective is fixed at the position indicated by dotted arrow in (b). The bottom objective is at the same position (dashed line) or at the position indicated by the black arrow (line). (d) Spatial map of the interference visibility. Experimental parameters are: $\lambda_{\text{laser}} = 860 $ nm, $\text{repetition rate} = 1 $ MHz, $\text{pulse duration} = 200 $ ns, $\text{power} = 35$ $ \mu$W, $V_{\text{Guard}} = 0.7 $ V, $V_{\text{Trap I }} = V_{\text{Trap II}} = 0.34 $ V, and $T_{\text{bath}} = 250 $ mK.}
\end{figure}
The challenge to operate a shift-interferometer with optical fibers in a cryogenic setting is to overcome slow shifts of the optical phase in each fiber, caused by temperature-induced changes of the fibers\textquoteright \ length and refractive index\cite{Lagakos1981}. One solution is to focus both objectives onto the same position on the sample and to utilize the interference of the transmitted and reflected laser light to compensate such phase shifts. The reflected laser light passes a retroreflector located on a piezo-stage in one arm of the interferometer [Fig. 1(a)]. Then, the interfering light is projected onto a photodiode. To this end, a dichroic mirror is utilized that reflects the laser light but transmits the PL of the sample (in the present case, only the PL of the dipolar excitons is transmitted). With the help of a proportional-integral-derivative (PID) feedback-loop, the current of the photodiode enables to control the position of the retroreflector and in turn, to stabilize the optical length of one interferometer arm with respect to the other. By inverting the sign of the PID loop, a discrete, relative length jump of $\lambda_{\text{laser}}/2$ can be achieved. This is done in the on-period of the pulsed laser. In its off-period, the interfering PL signal is measured [Fig. 1(c)]. The fast oscillations of the interferogram reflect the PID-controlled steps of the optical path difference with a step size of $\lambda_{\text{laser}}/2$. The slowly oscillating beating pattern $I_{\text{beating}}$ [dashed curve in Fig. 1(c)] is then proportional to the following envelope function 
\begin{equation}
I_{\text{beating}}\propto \pm \cos \left[2 \pi d_{\text{path}}\frac{\lambda_{\text{PL}}-\lambda_{\text{Laser}}}{\lambda_{\text{Laser}}\lambda_{\text{PL}}}\right]
\end{equation}

with $d_{path}$ the optical path distance between the two arms of the interferometer and $\lambda_{\text{PL}}$ the wavelength of the sample\textquoteright s PL. We deduce $\lambda_{\text{PL}} = 887 \pm 3$ nm from Fig. 1(c), which is the coherent PL of dipolar excitons in our sample. The latter is described in the following section. \newline
\begin{figure}
\centering
\includegraphics[width=8.5cm]{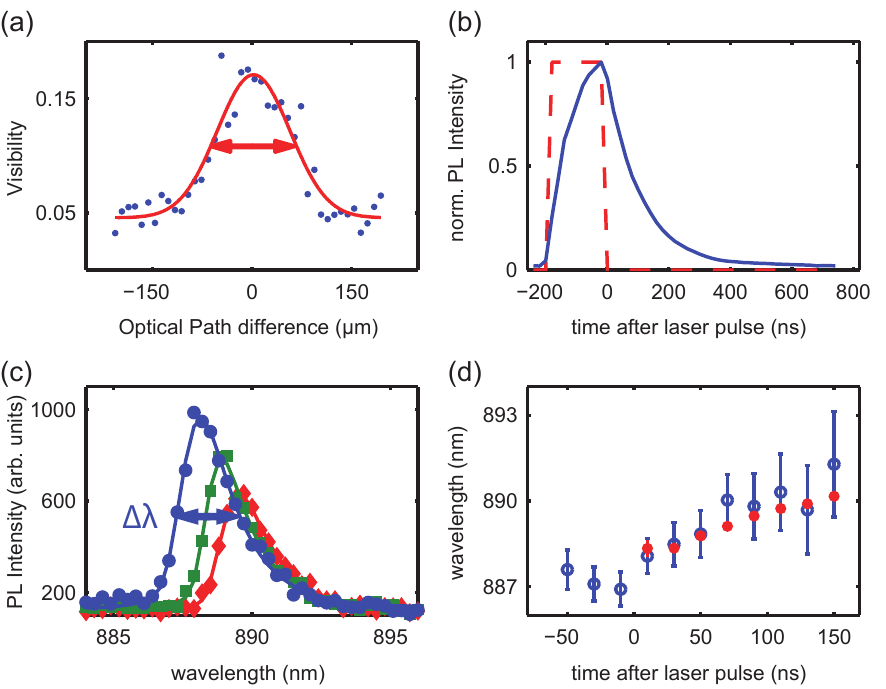}
\caption{\label{Fig3} Color online (a) Interference visibility vs. optical path difference (dots) and Gaussian fit. (b) Time lapse of the laser pulse (dashed curve) and spectrally integrated, normalized PL intensity (continuous curve). (c) PL spectra measured for different times after the laser pulse (dots 20 ns, squares 60 ns, diamonds 100 ns). (d) PL wavelength extracted from the spectra (dots) in (c) and from interferograms as depicted in Fig. 1(c) in combination with eq. (1) (open dots).}
\end{figure}
The sample consists of two 7 nm thick $\text{In}_{0.11}\text{Ga}_{0.89}\text{As}$ quantum wells, separated by a 10 nm thick GaAs barrier, embedded between a n-doped GaAs back contact and 6 nm thick, semitransparent titanium top gates defined by e-beam lithography\cite{Schinner2011,Schinner2013_1}. Fig. 2(a) depicts a scanning electron microscope image of the top gates of the sample. By applying an electric field between these gates and the back contact, an electrostatic trap can be formed for dipolar excitons within the double quantum well \cite{Rapaport2005,Chen2006,Hammack2006,Schinner2011,Rapaport2006,Remeika2012}. The trap behavior is verified by two-dimensional PL maps, where a signal is only recorded at the area of the gates trap I and trap II [Fig. 2(b)]. The excitons are excited through the top objective at the center of trap I (dotted arrow). The measured PL is the sum of the intensities in the two arms of the interferometer, while the bottom objective is laterally scanned. To avoid heating effects within the heterostructure, $\lambda_{\text{laser}} = 860$ nm is chosen to be slightly below the PL wavelength of the direct excitons within each of the two quantum wells\cite{Schinner2011}. In addition, the PL of the dipolar excitons is measured during the off-period of the pulsed laser. We find that the PL signal is slightly reduced at the position where the excitons are excited (dotted arrow). The reduction is persistent for a time scale exceeding the lifetime of the dipolar excitons [data not shown]. Therefore, we attribute the reduction to the excitation of mid-gap states within the field-effect structure, causing an additional local electric field at the excitation spot\cite{Alloing2013}. \newline
Fig. 2(c) shows two interferograms of the PL, where we make use of the thermal drift of the optical phase in the fibers. Without PID-stabilization, the thermal drifts cause a phase difference between the two arms of the interferometer. Whenever the optical phases of the PL in the different arms are correlated, the difference leads to a continuously varying PL intensity on the photodetector. Again, both objectives are focused on the same position indicated by the dotted arrow in Fig. 2(b). Within the measurement time of 9 s, thermal drifts cause a monotonous variation from constructive to destructive interference [dashed curve in Fig. 2(c)]. When the bottom objective is positioned at a different location [arrow in Fig. 2(b)] than the top objective [dotted arrow], the interference fluctuations of the PL signal disappear [line in Fig. 2(c)], although the drifts of the optical phase are still present. The described method allows to spatially map the coherence visibility of the PL without a PID-loop. For the data in Fig. 2(d), the top objective is fixed in the center of trap I and the bottom objective is scanned across the sample. The corresponding map shows the interference visibility that is calculated for each pixel as the quotient of the standard deviation and the mean value of the PL signal with an overall maximum of $\sim0.15$. When we apply the visibility as defined in ref. \cite{High2012_2} to the data in Fig. 2(d), we find $\sim0.48$, which is consistent with earlier reports \cite{Alloing2014}. We note that the PID-controlled interferometer, as described in the context of Fig. 1(c), only works for the situation when the laser light is transmitted from one objective to the other. \newline
Despite the versatility of our confocal microscope to spatially scan the two-dimensional coherence map of our sample [Fig. 2(d)], we observe a coherent signal only when both objectives focus on the same spot on the sample. A comparison between the two maps in Figs. 2(b) and 2(d) reveals that the interference visibility is independent of the intensity and fluctuations of the excitonic PL. Fitting a Gaussian to the interference pattern in Fig. 2(d) gives a FWHM of $1.2 \pm 0.3$ $\mu\text{m}$. This FWHM does not exceed the FWHM of the determined PSF of the two objectives. Accordingly, no spatial coherence of the excitonic emission is observed. This holds although our measurements are performed at a temperature of 250 mK, where a condensed phase of dipolar excitons should exist according to recent literature\cite{High2012_2,Shilo2013,Alloing2014}. Our results are consistent with the interpretation that the detected coherence of the excitonic emission stems from a temporal coherence. In the following, we discuss further experiments corroborating this interpretation. \newline
In the case that both objectives focus on the same position, the temporal coherence length of indirect excitons can be measured by increasing the optical path difference between the two arms of the interferometer [Fig. 3(a)]. For each data point, the temperature induced variations of the optical paths in the fibers are used to measure the interference visibility without PID-stabilization as described above. We detect a decreasing visibility for an increasing optical path difference centered around zero path difference. This observation is consistent with a temporal coherence. Fitting the measured data with a Gaussian gives a FWHM of $129 \pm 21 \mu\text{m}$ [arrow in Fig. 3(a)]. We note that the excitonic PL is only measured in the off-period of the laser. Hereby, we ensure that we can neglect a laser-induced coherence. For an optical path difference exceeding $\sim150 $ $\mu$m, the temporal coherence is lost. Therefore, we interpret the corresponding signal background of $\sim0.05$ to stem from fluctuations of the overall optical set-up. In our interpretation, this applies also to Fig. 2(d). \newline
The continuous curve in Fig. 3(b) shows the time-dependence of the normalized, spectrally integrated PL with a fitted lifetime of $119 \pm 5$ ns. Fig. 3(c) depicts the corresponding PL spectra for 20 ns, 60 ns, and 100 ns after the laser is switched off. The asymmetry of the PL spectra is discussed in refs.\cite{Schinner2013_2,Stern2014}. The data are fitted with an asymmetric pseudo-Voigt function giving a spectral width of $\Delta\lambda = 2.28 \pm 0.13$ nm for 20 ns\cite{Stancik2008}. In turn, a corresponding temporal coherence length can be calculated as

\begin{equation}
l_c= \frac{\lambda^2}{n \Delta\lambda}=96 \pm 6 \mu\text{m}
\end{equation}
where $n$ is the refractive index. This result is in good agreement with the temporal coherence length obtained from Fig. 3(a). In particular, the time-dependent spectra in Fig. 3(c) are measured for an acquisition period of 20 ns, which leads to somewhat broadened spectra. This explains the small deviation between the calculated coherence length in eq.(2) and the measured FWHM of the interferogram in Fig. 3(a). We note that the corresponding coherence time in the order of picoseconds is limited by the spectral diffusion corresponding to the observed PL linewidth\cite{Snoke2011}. The coherence time is well below the PL lifetime. \newline
With an increasing time after the laser pulse, the spectrally integrated PL decreases, and the PL spectra shift to lower energies [Fig. 3(c)]. The latter reflects a decrease of the blueshift of the excitonic PL caused by the screening of the electric field by the dipolar excitons. According to refs. [\cite{Butov1999,Schindler2008,Laikhtman2009,Ivanov2010}], the density of dipolar excitons in our trap can be calculated to be about $5 \cdot 10^{10} $ $\text{cm}^{-2}$ immediately after the laser pulse. 
With an exciton density well above $10^{10} $ $\text{cm}^{-2}$ and a temperature in the sub-Kelvin regime, the thermal de Broglie wavelength of the excitons exceeds the inter-excitonic distance \cite{Schinner2013_1}. Therefore, one would expect a condensation to the quantum mechanical ground state\cite{Laikhtman2009,Snoke2011}. In a theoretical work, it was predicted that the quantum mechanical ground state is formed out of excitons which do not couple to light\cite{Combescot2007}, and the condensate is therefore ’dark’. This might explain the absence of spatial coherence in the excitonic PL. Further causes could be  optically induced charge fluctuations and an insufficient visibility sensitivity. Recently, it also has been reported that due to a carrier exchange with dark condensed excitons, bright excitons should acquire a coherent phase\cite{Combescot2012}. In turn, a spatial coherence should emerge. We have to state that for our samples, we do not see any indication for such a ’gray’ condensate in the present data. \newline
Finally, we demonstrate that the shift of the PL wavelength for an increasing time after the laser pulse can also be read-out with the confocal interferometer operated with the PID-feedback. For each time interval after the laser pulse, an interferogram is recorded as in Fig. 1(c). It is then fitted with eq. (1) to extract $\lambda_{\text{PL}}$ [open dots in Fig. 3(d)]. The obtained values are consistent with the PL maxima achieved from fitted spectra such as in Fig. 3(c) [dots in Fig. 3(d)]. This consistency demonstrates that the confocal interferometer indeed senses the coherent emission of the dipolar excitons. We note that the PID-controlled interferometer can also be operated with a second laser with an energy lower than the photon energy of the dipolar excitons [see optional stabilization laser in Fig. 1(a)]. However, we  again observe a coherent signal only when both objectives focus on the same spot on the sample with an equivalent signal and noise (data not shown). \newline
In conclusion, we built a shift-interferometer based on a cryogenic confocal microscope. We demonstrate that with this setup, it is possible to map the coherence of trapped dipolar excitons with a high spatial resolution. We observe that the spatial spread of the visibility is consistent with the point spread function of the optical setup with two objectives. With this, we conclude that the excitonic emission exhibits only a temporal but no spatial coherence even at the lowest temperature of 250 mK.
\begin{acknowledgments}
We thank S. Stapfner, T. Faust, J. Rieger, S. Dietl, A. H\"ogele and S. Manus for technical discussion and the DFG under Project No. Ko 416/17, LMUexcellent, the BMBF - Q.com-H  16KIS0109, DFH/UFA  CDFA-05-06, and Mercur  Pr-2013-0001 for financial support.

\end{acknowledgments}
\bibliographystyle{apsrev4-1} 
\end{document}